\documentclass[journal]{IEEEtran}

\usepackage{ifpdf}
\ifpdf
  \pdfoutput=1\relax                   
  \usepackage{graphicx}                
  \DeclareGraphicsExtensions{.pdf,.png,.jpg,.jpeg} 
\else
  \usepackage{graphicx}                
  \DeclareGraphicsExtensions{.eps}     
\fi

\graphicspath{{figures/}{pictures/}{images/}{./}} 

\usepackage{microtype}                 
\PassOptionsToPackage{warn}{textcomp}  
\usepackage{textcomp}                  
\usepackage{mathptmx}                  
\usepackage{times}                     
         
\usepackage{cite}                      
\usepackage{tabu}                      
\usepackage{booktabs}                  
\usepackage{amsmath}
\usepackage{rotating}
\usepackage{wasysym}
\usepackage{url}
\usepackage{hyperref}
\usepackage{subfigure}
\usepackage{xcolor, colortbl}
\usepackage[binary-units=true]{siunitx}
\DeclareRobustCommand{\bigO}{
  \text{\usefont{OMS}{cmsy}{m}{n}O}
}

\title{Voreen - An Open-source Framework for Interactive Visualization and Processing of Large Volume Data}

\author{Dominik Drees$^\star$, Simon Leistikow$^\star$, Xiaoyi Jiang,
and Lars Linsen,
      \thanks{D.\ Drees, S.\ Leistikow, X.\ Jiang and L.\ Linsen are with the Faculty of Mathematics and Computer Science, University of Münster, Münster, Germany.}
      \thanks{$^\star$ These authors contributed equally to this work.}

}

\begin{document}

\maketitle

\begin{abstract}

Technological advances for measuring or simulating volume data have led to large data sizes in many research areas such as biology, medicine, physics, and geoscience.
Here, large data can refer to individual data sets with high spatial and/or temporal resolution as well as collections of data sets in the sense of cohorts or ensembles.
Therefore, general-purpose and customizable volume visualization and processing systems have to provide out-of-core mechanisms that allow for handling and analyzing such data.
Voreen is an open-source rapid-prototyping framework that was originally designed to quickly create custom visualization applications for volumetric imaging data using the meanwhile quite common data flow graph paradigm.
In recent years, Voreen has been used in various interdisciplinary research projects with an increasing demand for large data processing capabilities without relying on cluster compute resources.
In its latest release, Voreen has thus been extended by  out-of-core techniques for processing and visualization of volume data with very high spatial resolution as well as collections of volume data sets including spatio-temporal multi-field simulation ensembles.
In this paper we compare state-of-the-art volume processing and visualization systems and conclude that Voreen is the first system combining out-of-core processing and rendering capabilities for large volume data on consumer hardware with features important for interdisciplinary research.
We describe how Voreen achieves these goals and show-case its use, performance, and capability to support interdisciplinary research by presenting typical workflows within two large volume data case studies. 
\end{abstract}

\section{Introduction}
Voreen is a rapid-prototyping framework~\cite{meyer2009voreen} originally developed for in-core rendering of individual volume data sets  and related research.
With the technological advances in data generation, new challenges arose that were addressed during the continued development of Voreen.
New and improved imaging technologies, in particular, in the fields of biology and medicine~\cite{kirst2020mapping,zhao2020Cellular}, led to steadily rising spatial resolutions and nowadays are capable of producing \textit{extremely large single-volume data sets}, where {extremely large} today refers to data sets in the order of hundreds of Gigabytes and beyond.
Dynamic imaging techniques that produce \textit{time-varying volumes} are also employed.
Traditional approaches considered one data set at a time. With the upcoming of personalized medicine and epidemiological studies such as national \textit{cohorts}, there is an increasing need of handling and analyzing multiple data sets simultaneously.
Measured data can also be complemented by simulated data to overcome measuring obstacles.
As basically all simulations depend heavily on the choice of initial configurations and parameter settings, it is a common approach to generate \textit{simulation ensembles} with multiple simulation runs that capture variabilities, dependencies, and uncertainties.
Thus, we are facing the challenge of processing and visualizing 5D data, where the volume data vary over time and over parameter settings.

Since its original release and publication~\cite{meyer2009voreen}, Voreen has been extended to handle and analyze such large data, which has now been put into the open-source framework within this year's release.

Voreen's features with regard to handling large out-of-core volumes range from generally applicable and commonly applied processing steps and visualization functionality to more specialized, application-specific components. Many features are commonly applied (sequentially) to individual volumes.
They include basic, but efficient image processing/filtering, connected-component-analysis\cite{isenburg2009streaming}, rendering with support for multi-channel volumes with channel-shift~\cite{brix2014visualization}, efficient interactive segmentation~\cite{drees2021hierarchical} as well as vessel network analysis and quantification tools~\cite{drees2021scalable}.

When it comes to analyzing cohorts or ensembles, Voreen provides high-level visualizations (such as Multi-run Similarity Plots~\cite{fofonov2019projected}, Parallel Coordinates, time histograms, and variance calculation) as well as low-level operations and routines (such as filtering and transparent out-of-core-processing) for spatio-temporal multi-field or multi-channel ensemble data sets. 
When combined with other existing features, these methods allow developers to implement all sorts of analysis workflows including visual encodings that operate at multiple levels of detail and abstraction.

One aspect of Voreen's mission has always been to facilitate \textit{interdisciplinary research}. The development of the above-mentioned features is driven by ongoing interdisciplinary research between computer science (volume processing/visualization/analysis) and application domains such as biology and medicine, material science, physics and geosciences, etc.
With respect to large data, there are new functionalities that need to be added to existing features. Within many years of conducting interdisciplinary research we identified requirements that can be summarized as \textit{algorithmic extension option}, \textit{bulk processing}, \textit{headless execution}, \textit{task-specific applications}, and \textit{animation support} which will be further elaborated in \autoref{s:survey}.
Implementing the features is enabled by Voreen's architecture:

The core of Voreen's processing pipeline is the \textit{data flow graph}.
On the one hand, it enables developers to rapidly develop and test new visualization and data processing tools as new nodes in the graph.
On the other hand, these new features can easily be integrated with existing components to build complex applications with low effort.
The \textit{application mode} reduces the powerful data flow graph user interface (used by developers) to a task-specific view that abstracts from non-critical details (used by application scientists). 
Results of the analysis process can be reported and disseminated with the \textit{animation editor} and video export functionality.
Moreover, the same components and data flow network definitions can be used for complex quantitative evaluation via a \textit{headless interface} and using the integrated python scripting capabilities, including \textit{bulk processing}.
The value and contribution to the scientific community of Voreen and, in particular, its recent advancements is evidenced by the fact that it is used as a foundation in many research projects~\cite{athawale2020direct, leistikow2020interactive, drees2021hierarchical, johansson2018colored} and is widely used for its existing functionality in biomedical \cite{benz2019low,rene2017,ingendoh2018time} and broader natural science~\cite{landell2019material} research.

In this paper we present Voreen's new large volume data processing and visualization capabilities as well as its features that are essential for interdisciplinary work:
For this, we first review available commercial and open-source volume processing and visualization software packages in \autoref{s:relatedwork} and then compare Voreen to them with a focus on large data capabilities and interdisciplinary research support in \autoref{s:survey}.
Voreen's functionality with respect to interdisciplinary research support (\autoref{s:interdisciplinaryresearch}), large volume processing and visualization (\autoref{s:ultralargevolumes}), and ensemble analysis (\autoref{s:visualensembleanalysis}) is detailed afterwards.

For both large volumes and ensembles we provide a case study to demonstrate how Voreen can be used to support interdisciplinary research (\autoref{s:evaluation}).
Finally, we discuss technical limitations (\autoref{s:limitations}) and conclude with an outlook on further development of the framework (see~\autoref{s:conclusion}).

\smallskip
\noindent
Our main contributions of this paper can be summarized as presenting
\begin{itemize}
    \item an open-source framework that supports processing and interactive visualization of large volume data, i.e. out-of-core volumes and ensembles (executable files for both linux and windows, example projects, and sources can be obtained freely  at \url{https://www.uni-muenster.de/Voreen/}),
    \item an in-detail comparison to other software packages with respect to out-of-core processing and rendering of large single- and multi-volume data as well as support of interdisciplinary research, and
    \item case studies for both large volumes and ensemble data sets to demonstrate how Voreen can be used to support interdisciplinary research.
\end{itemize}

\section{Related Work}
\label{s:relatedwork}

Over the last decades, many visualization and data processing frameworks for volumetric imaging data have emerged.
This includes commercial applications~\cite{amira,mevislab,avs,imaris} as well as free and open source software~\cite{jonsson2019inviwo,ahrens2005paraview,bavoil2005vistrails,jiang2016advanced,clyne2007interactive,schindelin2012fiji,gmd-8-2329-2015,fogal2010tuvok}.

\smallskip
\noindent
\textbf{Libraries.}
Foundation to many visualization tools is The Visualization Toolkit (VTK)~\cite{vtk}, a C++ framework which provides many broadly applicable data structures and algorithms, not only for imaging data. VTK-m~\cite{moreland2016vtk} is being developed to make better use of parallel architectures than VTK. The Insight Toolkit (ITK)~\cite{itk} is another C++ framework that provides even more data structures and algorithms, mainly for the segmentation and registration of imaging data, but does not contain any visualization component.

\smallskip
\noindent
\textbf{Commercial tools.}
Amira~\cite{stalling2005amira,amira} is a commercial tool designed and used for processing, rendering, and visual analysis of volumetric imaging data of all sorts of modalities from many domains such as biology, material sciences, and astrophysics.
The main features are part of a base version, whereas more specialized features, such as support for data sets larger than \SI{8}{\giga\byte} in size, are structured into modules that need to be purchased separately.

MeVisLab~\cite{mevislab} is a commercial (but free to use for non-commercial purposes) framework for the development of clinical application prototypes.
It uses the network metaphor for defining data processing pipelines, allows for algorithmic extension via C++, and exposes a python and javascript API.
Custom user interfaces can be created using the MeVisLab Definition Language (MDL).

Imaris~\cite{imaris} is a commercial software for microscopy image analysis, offering a number of different features depending on the package that was purchased.
Depending on the package, it has some support for large volume rendering and processing, as well as batch processing.

AVS Express~\cite{avs} is a commercial tool for creating visualization applications using predefined modules.
It seems to be the basis of a consulting operation and hence its exact feature set is hard to determine and therefore not discussed in \autoref{s:survey}.
It allows for application in cloud/hpc-environments, but it does not advertise support for ensembles or large volumes.

\smallskip
\noindent
\textbf{Free and open-source tools.}
Inviwo~\cite{jonsson2019inviwo} is a recently published open-source, multi-platform visualization system aimed at visualization developers with a focus on debuggability of visualization pipelines.
As such, it utilizes the data flow graph and provides support for extensions via custom processors or shaders.
It also supports the creation of task-specific applications.
However, the publicly released version currently lacks support for large volume or ensemble processing and rendering.

VisTrails~\cite{bavoil2005vistrails} is another open-source visualization system using the popular data flow network metaphor.
Its main feature is scientific workflow and provenance management.
It does not support large volumes or ensembles and has been unmaintained since 2017.

ParaView~\cite{ahrens2005paraview} is an open-source, multi-platform data analysis and visualization application with large data visualization as the mission target.
In contrast to our work, however, this is done by distributing the data set and computational load on a distributed compute cluster.
It allows for building visualization pipelines using a tree view and has support for batch processing.

Tomviz~\cite{jiang2016advanced} is an extension to Paraview with a focus on processing, visualization, and analysis of 3D tomographic data.
It inherits Paraviews limitations with regard to out-of-core data.

VisIt~\cite{childs2012visit} is another (primarily) distributed open-source visualization system, with a focus on application to simulation results.
It supports visualization and some processing of a wide range of data types, including single volumes and ensembles.
Like Paraview, however, support for large data sets is achieved by operating in a cluster environment.

ImageVis3D is a visualization application for out-of-core volume data sets. It is based on Tuvok~\cite{fogal2010tuvok}.

VAPOR~\cite{clyne2007interactive} is an open-source visualization system for weather phenomena with a set of predefined renderers.
Due to its narrow focus it only supports loading of NetCDF, WRF-ARW and MPAS files.

Fiji~\cite{schindelin2012fiji} is a multi-platform, open-source distribution of the image processing library ImageJ~\cite{rueden2017imagej2} with a graphical user interface, an integrated package manager, and a large number of community-provided plugins.
Its main focus is to assist research in life sciences and, in particular, 2D image processing, but it also provides 3D functionality (depending on the plugin).
Handling of data larger than main memory is possible in principle (an example is the BigDataViewer~\cite{pietzsch2015bigdataviewer}), but not supported in the general case.

Met.3D~\cite{gmd-8-2329-2015} is an open-source tool that focuses on weather forecasting data, i.e., it provides related tools such as interactive visual analysis of probability volumes and supports data aggregations such as ensemble mean and variance. The tool assumes, however, that single time step volumes fit into GPU memory.

Seg3D~\cite{cibc2016seg3d} is an open-source volume processing and segmentation tool that includes visualization features for this purpose.
It is restricted to handling volume files that fit into main memory.

MegaMol~\cite{gralka2019megamol} is a molecule rendering framework.
As such, it is focused on visualization of point-based data sets and hence not considered further here.

\newcommand{\tl}[1] {\begin{turn}{90}#1\end{turn}}
\newcommand{\fullCapable}{\newmoon}
\newcommand{\partCapable}{\fullmoon}
\newcommand{\noneCapable}{}
\newcommand{\todoCapable}{?}
\begin{table*}[]
    \centering

    \begin{tabular}{clllllllllllll}

                                          & \tl{Amira}     & \tl{Fiji}    &\tl{ImageVis3D} & \tl{Imaris}    & \tl{Inviwo}  & \tl{Met.3D}  & \tl{MeVisLab}& \tl{ParaView}& \tl{VAPOR}  & \tl{VisIt}   &\tl{VisTrails} & \tl{Voreen}  & \tl{tomviz}  \\
        \hline

        Large volume processing/rendering & \fullCapable * & \partCapable & \partCapable   & \fullCapable * & \noneCapable & \noneCapable & \partCapable & \partCapable & \partCapable& \partCapable & \noneCapable  & \fullCapable & \noneCapable \\
        \hline
        Ensemble processing/visualization & \partCapable   & \noneCapable & \noneCapable   & \noneCapable   & \noneCapable & \fullCapable & \todoCapable & \partCapable & \partCapable& \fullCapable & \noneCapable  & \fullCapable & \noneCapable \\
        \hline
        Algorithmic extension option      & \fullCapable * & \fullCapable & \partCapable   & \noneCapable   & \fullCapable & \partCapable & \fullCapable & \fullCapable & \fullCapable& \fullCapable & \fullCapable  & \fullCapable & \fullCapable \\
        \hline
        Bulk processing                   & \fullCapable   & \fullCapable & \fullCapable   & \fullCapable * & \fullCapable & \fullCapable & \fullCapable & \fullCapable & \noneCapable& \fullCapable & \fullCapable  & \fullCapable & \fullCapable \\
        \hline
        Headless execution                & \fullCapable   & \fullCapable & \partCapable   & \noneCapable   & \partCapable & \noneCapable & \noneCapable & \fullCapable & \noneCapable& \fullCapable & \fullCapable  & \fullCapable & \noneCapable \\
        \hline
        Task-specifc applications         & \noneCapable   & \partCapable & \partCapable   & \noneCapable   & \fullCapable & \noneCapable & \fullCapable & \partCapable & \noneCapable& \partCapable & \noneCapable  & \fullCapable & \noneCapable \\
        \hline
        Animation support                 & \fullCapable   & \partCapable & \partCapable   & \fullCapable   & \fullCapable & \partCapable & \partCapable & \fullCapable & \partCapable& \fullCapable & \fullCapable  & \fullCapable & \fullCapable \\
        \hline
        Free and open source software     & \noneCapable   & \fullCapable & \fullCapable   & \noneCapable   & \fullCapable & \fullCapable & \noneCapable & \fullCapable & \fullCapable& \fullCapable & \fullCapable  & \fullCapable & \fullCapable \\
    \end{tabular}
    \caption{
        Feature overview of applications that allow the definition of custom visualization/data processing pipelines and/or support processing or rendering of out-of-core data.
        (\fullCapable: fully capable, \partCapable: partially capable, *: Depending on available extensions/modules.)
    }
    \label{tab:feature-overview}
\end{table*}

\section{Software Package Review}
\label{s:survey}

In this section we will review frameworks with respect to their support for interdisciplinary research and support for large volume and ensemble data.
For each aspect, we restrict discussion to systems that do (to the best of our knowledge) support this aspect at least in some respect.
We will restrict our comparison to applications (rather than libraries such as itk or vtk(-m) which are in part used by these applications), which allow the definition of custom visualization/data processing pipelines and/or support processing or rendering of out-of-core data.
For an overview of the systems and their supported features, we refer to \autoref{tab:feature-overview}.

\paragraph{Out-of-core Processing and Rendering}

To process out-of-core data, it is mandatory to split the data into chunks that can be processed in main memory or in VRAM, when it comes to rendering or other hardware-accelerated calculations.
In this regard, two different approaches are implemented in the reviewed software packages. Either the data chunks are processed mostly sequentially on the same system or the data gets distributed to multiple computing nodes, e.g. over the network, and the results are streamed back to a client to be further processed or visualized.
The distributing approach has certain overhead as consequence that might render it unsuitable for an interactive visualization.
It is therefore mainly used for data processing or to generate non-interactive image or video data.
This approach is implemented in the ParaView and VisIt software packages.

Since not everyone has access to a compute cluster or an interactive analysis might be required to understand the data, other software packages handle large volume data differently, e.g. by down-sampling or by the use of sophisticated chunking/streaming approaches. To achieve this, Amira makes use of so-called \textit{Large Data Access} (LDA) files whereas Voreen uses an octree datastructure for rendering and some types of processing, described in more detail in \autoref{s:ultralargevolumes}.
Amira advertises processing of ``any modality, at any scale, of any size'' and mentions in particular filtering operations, artifact removal, and image stack alignment.
Support for rendering of data sets larger than 8GB, however, requires access to the ``XLVolume'' extension.
Similar to Voreen's approach, ImageVis3D first converts an input volume into a LOD representation, which allows for rendering out-of-core data sets.
However, it is solely focused on rendering and, as such, does not have volume processing capabilities.
For rendering large volume data sets in VAPOR, users can convert their data into \textit{Vapor Data Collection (vdc)} files using provided command line applications.
Data in these formats can be rendered using different levels of compression, allowing interactive visualization at high compression levels, and offline high-fidelity rendering.
However, the user is responsible for selecting a level of detail, that does not exceed the memory capacity.
Streaming rendering and processing of a single data set does not appear to be supported.
Imaris supports rendering of out-of-core data sets~\cite{brix2014visualization} and offers image stitching capabilities for terabyte sized volumes (depending on access to the ``Imaris Stitcher'' package).
Out-of-core volumes can be rendered in Fiji as slices using the BigDataViewer-plugin~\cite{pietzsch2015bigdataviewer}, but we are not aware of a plugin supporting direct volume rendering.
The Inviwo framework~\cite{jonsson2019inviwo} claims to have rendering support for out-of-core data. However, that feature has not yet been released to the public and, thus, could not be tested.
In MeVisLab, loading and processing of out-of-core volumes is possible when using the DICOM format.
Volume files of others formats are loaded by ITK image readers and therefore are restricted by their size and the main memory of the host system.
3D rendering of out-of-core data sets is possible. However, it makes excessive use of hardware acceleration and freezes the rest of the application while processing a new full quality image (tested on a 13GB volume).
Trying to create and render a larger, yet still moderately sized volume (90GB) crashed the application.

Generally, a problem for loading large data sets immediately arises for many of the discussed applications, because they rely on vtk for loading the input files.
The development of vtk-m may improve the situation in the future, but it should be noted that proper handling of large volumes is required in all parts of the application, including processing steps, which may, for example, be required to save intermediate results to disk before rendering.

\paragraph{Ensemble Processing and Visualization}

Most of the tools support time series in some way, e.g. by rendering individual volumes of the sequence. However, only few can handle time series that do not fit into main memory, i.e., by loading and releasing required time steps on demand (Amira, Met.3D, MeVisLab, ParaView, VAPOR, VisIt, and Voreen).
An Ensemble can be understood as a set of such time series. Providing appropriate loading mechanisms and data representations for further operations such as data aggregation and abstraction is essential to create visualizations that are able to convey all facets of the data. 
To the best of our knowledge, only Met.3D, VisIt, and Voreen provide the tools needed for data aggregation and abstraction.
In VisIt, ensembles are interpreted as databases on which queries can be executed, however, no generally applicable ensemble visualization method or data structure is provided. Similarly, most other tools are able to perform some sort of aggregation by making use of their bulk-processing capabilities, such as Python scripts. Met.3D provides visualizations such as probability volumes for weather forecasting but is limited to data from this particular domain.
Voreen, instead, provides data structures and visualizations generalized to any type of imaging data.

\paragraph{Support for Interdisciplinary Research}

Supporting interdisciplinary research has many aspects which we have grouped into the identified requirements.

\smallskip
\noindent
\textbf{Algorithmic extension option.}
For a software framework to be used for research at the interface between computer science and other disciplines, a way to extend the application with custom algorithms is essential.
Without this aspect, the application is merely a tool for application of existing methods, and not a platform for the development of new techniques.
Amira (if the XPand extension was purchased) and MeVisLab allow users to extend the functionality with custom C++ modules.
All open-source frameworks -- due to their licensing -- implicitly allow algorithmic extension by arbitrary changes to the code.
This is made easier by providing abstractions such as the popular data flow network metaphor employed in Amira, MeVisLab, Inviwo, VisIt, VisTrails, and Voreen or the tree abstraction of ParaView and its extension tomviz.
Inviwo additionally supports to extend the functionality using python scripts and shaders and by hot-reloading parts of the processor network after recompilation.
VisIt, Paraview, Vistrails, and Fiji allow extensions via plugins/packages.
VAPOR has python support for data processing/generation, while tomviz can use python scripts for custom operators and file formats.
Voreen allows for the creation of custom modules and processors that are decoupled from the application core, but can use the provided data structures.
Processor functionality can also be implemented in python scripts.

\smallskip
\noindent
\textbf{Bulk processing.}
When a tool was used to develop a data processing pipeline, it is usually desirable to apply the same pipeline to a large number of similar input data sets in order to allow for a statistically significant quantitative evaluation.
Almost all surveyed applications with the exception of VAPOR provide some form of support for bulk processing, at least from within the application.
A popular method to allow bulk processing is to allow scripting via python (Amira, Paraview, MeVisLab, Inviwo, VisIt, tomviz, Fiji, and Voreen).
Amira additionally provides a tcl interface, MeVisLab can also be controlled via javascript, and Fiji allows for scripting in various languages.
ImageVis3D's behavior can be influenced via lua scripts.
VisTrails workflows can be processed in bulk using the VisTrails server application.
Met.3D has a batch mode that can be used to automatically render and save a time series dataset as a sequence of images.
Batch processing in Imaris requires the \textit{Imaris Batch} package.

\smallskip
\noindent
\textbf{Headless execution.}
As individual processing steps may take a long time to execute, it is often desirable to move the execution of expensive pipelines to a dedicated, more capable, often headless machine, i.e., a machine without a graphical user interface.
Paraview and VisIt allow distributed and thus implicitly headless execution on clusters.
Similarly, for VisTrails, its server application can be used to execute pipelines remotely.
Inviwo's functionality can be made available as a python package, thus theoretically allowing for headless execution, but we were unable to execute it in a headless environment due to it unconditionally trying to create an OpenGL context.
Similarly, the ImageVis3D source includes a BatchRenderer target that can be used to execute lua scripts, which, however, require the creation of a graphics context for rendering.
Fiji, in constrast, can be used to execute predefined scripts or macros in a fully headless environment.
Amira provides an option to start without a graphical user interface, but we were unable to verify if it truly works headlessly, since we do not have access to a licensed copy of the commercial software.
Voreen allows for the execution of previously defined pipelines (optionally in conjunction with a controlling python script) in a command line application.

\smallskip
\noindent
\textbf{Task-specific applications.} For the definition of visualization and processing pipelines, a powerful and flexible, but necessarily \textit{developer}-focused user interface is required and usually provided.
This can be done in many forms, e.g., by interactive manipulation of a data flow network, or by providing a library or scripting interface.
On the other hand, the \textit{application} of a defined pipeline has different needs for a user interface, which should instead be intuitive and focused on the specific task that the pipeline enables.
Ideally, a volume visualization and processing framework should provide a simple way to define and create such task-specific applications.
MeVisLab, Inviwo, and Voreen have support for the creation of custom user interfaces built into the application itself:
In MeVisLab custom GUIs are described using the MeVisLab Definition Language (MDL).
Inviwo allows for hiding the network representation and shows a selected list of changeable parameters in a separate pane.
Voreen's application mode hides \textit{all} elements not required for the specific task and only shows the canvas and a pane of selected properties that are grouped into custom categories.
ParaView, Inviwo, and Visit provide a library interface, which, although it requires significant development effort, also enables the creation of custom GUIs.
The same is true for ImageJ~\cite{rueden2017imagej2}, which is the basis of Fiji, as well as for Tuvok~\cite{fogal2010tuvok}, which is the backend for ImageVis3D.

\smallskip
\noindent
\textbf{Animation support.} While interactive visualization is an important aspect, e.g., for purpose of data exploration, for communication and sharing of the findings of the exploration process, it is also desirable to export the visualization results for use outside of the application.
In the simplest form, this can be done by saving snapshots as images, but a moving image, i.e., a video is often more engaging and has the potential to transmit more information to the viewer.
For smooth changes of parameters of the visualization, some form of built-in support animation is required.
Many of the discussed tools support some kind of animation and video export.
A popular approach (employed by Paraview, tomvis, Amira, and Inviwo) is the definition of key frames for selected properties and interpolation for intermediate values.
Similarly, Imaris and Voreen support the creation of global key frames, which define values for selected parameters of the visualization and allow for interpolation in-between.
Vistrails requires the user to first create a table of changeable parameters (via its ``explore'' feature) which can then be used to create an image sequence.
ImageVis3D and MeVisLab allow for the creation of animations via a fixed rotation of the camera around the object.
VAPOR and Met.3D enable animation in the time dimension of 4D data sets.
Fiji has no built-in support for animations, but snapshots of the current rendering can manually be converted into an image sequence within the application.

\medskip
As shown above and summarized in \autoref{tab:feature-overview}, none of the existing state-of-the-art systems except for Voreen provides the full feature set required to support interdisciplinary research on large volume data on commodity hardware.
In the following sections, we will detail how Voreen fills this gap.

\section{Support for Interdisciplinary Research}
\label{s:interdisciplinaryresearch}

One of Voreen's goals is to facilitate research at the boundary between computer science and other natural sciences as well as medicine.
This is achieved by exposing the core functionality of Voreen in different ways for different users and applications.
Moreover, dedicated features such as the video animation and export support are provided.

\subsection{Data Flow Graph Interface}
Like other tools, Voreen allows the ad-hoc definition of data processing and visualization pipelines by enabling the user to define and modify directed acyclic data flow graphs using a large number of predefined or custom \textit{processors} that act as nodes in the graph, read data (such as volumes, ensembles, images) from incoming edges, and produce data to be consumed by other, downstream processors.
Data inputs, e.g., loading volumes from disk and supporting a number of common file formats (dicom, tiff, hdf5, nifti), form the sources in the data flow graph, while data exporters (for whole data set or the results of analysis pipelines in the form of csv or json data) or a canvas (for immediate display to the user) act as the sinks.
The behavior of processors can be adjusted and customized using \textit{properties} that define numerical, categorical, boolean, or other parameters of the processing algorithm.
Properties of different processors may also be linked, which automatically synchronizes their values upon changes to either of them.

\textit{Workspaces} aggregate the data flow network as well as current property values and links of processors which can be stored and loaded as custom xml files.
As a result, workspaces can be used to define, store, edit and distribute data processing and visualization applications, using voreen as a framework for execution.

Either by composing a workspace from existing processors, or by writing custom processors, a user with data processing or visualization background is able to rapidly create and define domain specific applications that can then be used to facilitate experiments as well as data exploration, processing and evaluation.

\subsection{Application Mode}
\begin{figure}[t]
  \includegraphics[width=\columnwidth]{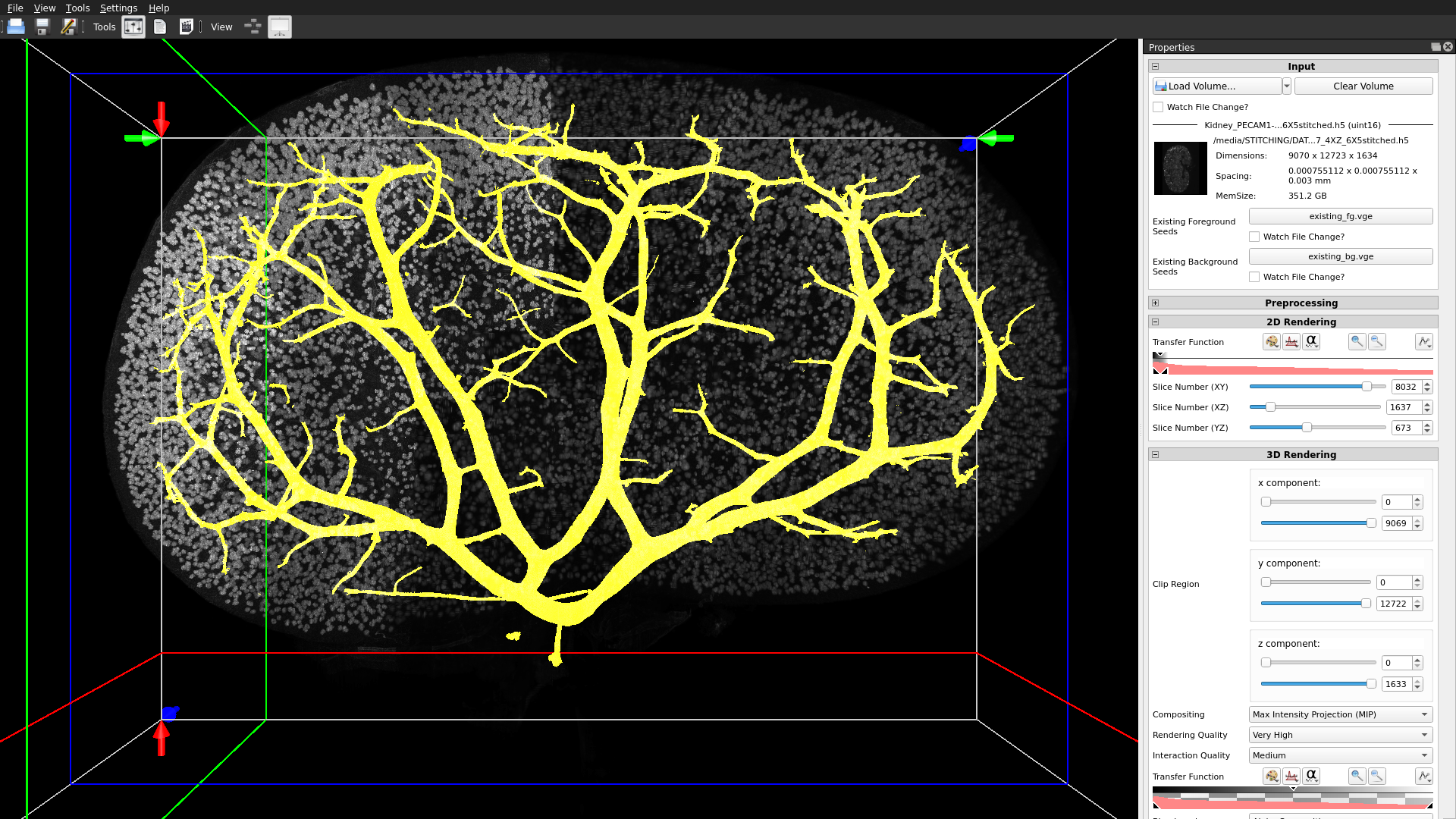}
  \caption{
    A screenshot of the Voreen application mode interface with a 3D raycasting rendering of the \SI{377}{\giga\byte} mouse kidney data set and the foreground segmentation of the arterial vessel tree created with the hierarchical random walker framework~\cite{drees2021hierarchical}.
    The canvas shows the maximized 3D view (three axis-aligned slice views for labeling are hidden) and the curated set of properties important for the current segmentation task in the panel on the right.
    The progressive rendering approach is visible by high-detail bricks already rendered on the left and low-fidelity rendering remaining in the right part of the image.
    \label{fig:progressive}
  }
\end{figure}

While the data flow graph-based user interface is powerful for both editing and using data processing and visualization pipelines, it may be daunting and too complex for domain experts without data processing background, which want to act as \textit{users}, but not \textit{authors/developers} of pipelines defined within the Voreen framework.
For this reason, Voreen includes an \textit{application mode} with a reduced user interface, which abstracts from the data flow graph and related editing options.
Instead, it only shows the rendering of a single designated canvas of the network, as well as selected processor properties in a single property pane on the screen.
Each property in the pane corresponds to a specific property of a processor in the data flow network and, thus, influences its behavior, and optionally those of linked properties.
For this view, properties can also be renamed and aggregated into groups (corresponding, e.g., to a specific stage in a processing pipeline) in order to provide a more application-focused context and terminology to the user.
Workspaces also include a description, which aims to guide an inexperienced user in using the domain-specific application.
\autoref{fig:progressive} shows an exemplary task-specific application for semi-automatic segmentation of out-of-core volumes.
Within the application, the user can interact with properties, e.g., the transfer function currently used to render the data set, as well as the main view, where in the example the 3D view was maximized by double clicking, thus hiding the three 2D views used for label definition.

The configuration of the application mode is also stored as part of the workspace and can be edited in a dedicated dialog.

Additionally, all core functionalities of Voreen are provided as a shared library, which enables the further development of completely independent applications, after prototyping its functionality using the application mode of Voreen.

\subsection{Headless Execution and Bulk Processing}
In addition to interactive applications, workspaces can also be used for the definition of \textit{offline}, headless data processing (and static visualization) pipelines.
For this, the voreen framework provides a command line application (\textit{voreentool}) which, when executed with a workspace as a parameter, constructs the processing network, and performs a single evaluation -- from loading data sources, via intermediate processors, to writing result files -- of the network.
Here, the behavior can be customized by overriding the saved property values with values specified as command line arguments.

For even more complex cases and interoperability with the rich python ecosystem, Voreen allows customization of the evaluation process using python scripts, where it exposes its core functionality as python functions and objects.
This way the user can define and change property values and control the execution of Voreen's event loop and network evaluation.

This procedure also enables bulk processing:
The user can loop over data sets of interest by loading each data set, (optionally) updating parameters, evaluating the network, and saving results.
Such a script for bulk processing can be used either in the visual environment or with the command line application.
Alternatively, for simple cases that do not require advanced control over the network execution, the command line application can be used for bulk processing on its own, i.e., by executing it separately for each data set and applying changes to property values via command line parameters.

Additionally, it is possible to integrate external python libraries by using the PythonProcessor in the network, for which the processing step is defined by a python script that can read incoming and write outgoing data.

It should be noted that a single workspace can serve both as an interactive application \textit{and} as the definition of an \textit{offline} processing pipeline.
Which role it assumes is only dependent on whether it is executed using the visual environment or the headless execution tool.

\subsection{Animation and Video Support}
For research outreach, presentation, or communication between collaborators, it is becoming increasingly important to provide a concise, but intuitive summary of goals, results, or  current state of a project. For this purpose, representative images or even better video data serve as a powerful tool to convey required information.

To support this, Voreen provides both property animation and video export functionality.

In the animation editor, properties selected for animation can be assigned specific values at key frames in the timeline.
Between key frames, the property values are determined by interpolation, for which various options exist (linear, spherical linear, and spline-based).
This allows for the definition of a consistent animation procedure, which can be applied, for example, to different data sets while ensuring that the results are comparable.
For a given animation, a sequence of frames can then be generated and exported either as a picture sequence or as a video.

\section{Processing and Rendering Large Volumes}
\label{s:ultralargevolumes}
Advances in microscopy technology allow the acquisition of single volume data sets that lead to hundreds of Gigabytes or even Terabytes.
They promise greater insights into biological processes to be revealed by research~\cite{kirst2020mapping,zhao2020Cellular}.
These data sizes, however, create a need for algorithms and methods that are able to process these data sets, as well as a framework which makes existing methods available for application.
Voreen aims to fulfill this need by providing a platform for the development of new methods and by including support for various existing methods of processing and rendering of single large, i.e., out-of-core volumes that are described below.
The methodology includes general approaches such as filtering, rendering, segmentation, and voxel-based quantification, as well as more specialized (but especially in the biomedical domain broadly applicable) methods for vessel network and cell cluster analysis.

For user comfort, all methods are implemented as an asynchronous process, which does not block the user interface and (when configured as such) automatically (re)starts when the input in the data flow graph becomes ready or is changed.

\subsection{Large Volume Data Access}
Voreen supports different volume data formats (e.g., ome.tiff, dicom, hdf5, nifti, and vvd) that allow for reading parts of the whole volume into memory, which is (obviously) essential for handling out-of-core volumes.
In these formats, the typical case is that the volume data is linearized into a stream of voxels, by convention in the axis order $zyx$, i.e., with the $z$-coordinate changing least frequently.
This storage scheme allows for reading individual xy-slices of the whole volume.
Hence, this type of access is assumed as a common interface for operations that directly operate on a (potentially external) volume, i.e., on the disk representation of a volume.
If a small volume is only present in RAM, it supports this operation as well and can therefore be used as an input for the operations described below, as well.

However, especially for interactive operation, this may not always be the best representation and access method.
In cases where the (computational or visual) focus lies in a specific region (not necessarily within one slice) of the volume, an octree data structure (originally introduced by LaMar~\cite{lamar2000multiresolution} and Weiler~\cite{weiler2000level}) is beneficial and can be created by converting a disk representation  in a highly optimized, parallel preprocessing step.
This conversion takes only a few minutes for data sets of hundreds of Gigabytes and results are additionally cached for the most recently used volumes.

\subsection{Filtering}
\label{sec:filtering}
Large intermediate results in the processing graph are (by necessity) written back to disk (potentially compressed, either in linear form or in an octree representation) by each processor.

\begin{figure}
    \centering
    \includegraphics[width=\columnwidth]{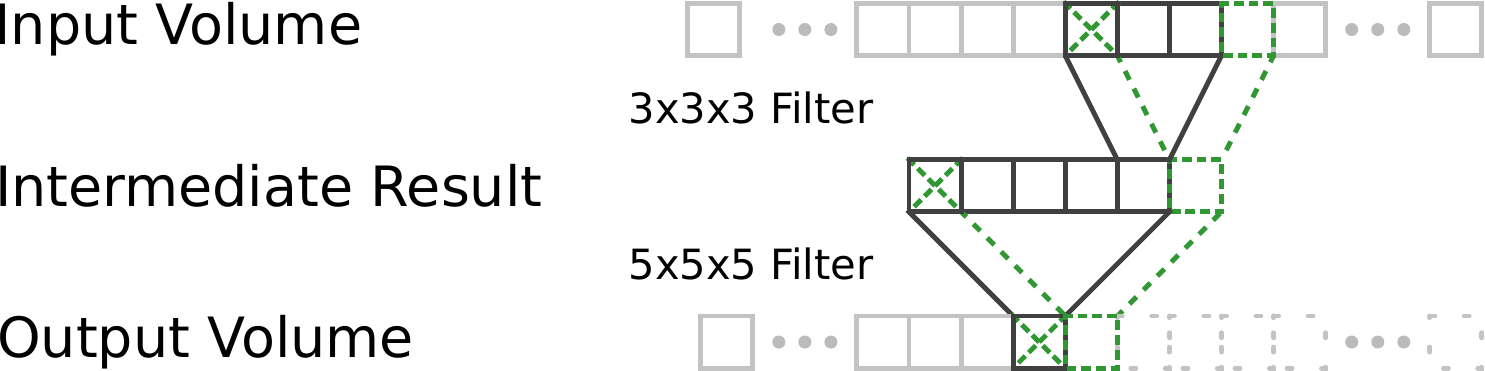}
    \caption{Illustration of the operation of a filter stack that does not read/write intermediate results to disk.
    By keeping a number of slices equal to the $z$-depth of the filter in memory and updating this set one slice at a time, no additional computations are required.}
    \label{fig:filter_stack}
\end{figure}

Often processing pipelines include a series of basic operations, for example, as a preprocessing step.
As these operations are typically computationally simple, large percentage of wall-clock processing time can be attributed to reading from or writing to disk.
For linear parts of the pipeline where intermediate results are not required, this materialization can be avoided:
All filters operate on a set of $k$ input slices $\{I_{i+1}, \ldots, I_{i+k}\}$ to produce an output slice $O_k$.
Notably, the set of input slices for $k$ and $k+1$ given by $\{I_{i+1}, \ldots, I_{i+k}\} \cap \{I_{i+2}, \ldots, I_{i+k+1}\} = \{I_{i+2}, \ldots, I_{i+k}\}$ overlap in such a way, that after generation of $O_k$, in order to produce $O_{k+1}$ only a single input slice (namely $I_{i+k+1}$) must be loaded from disk or generated by another filter.
At the same time, $I_{i+1}$ can be released.
The sequence of filters to be applied to an input volume forms a stack (with the first filter to be applied on the top), on which the filters are advanced a slice at a time (from top to bottom), in order to generate the next output slice that is materialized and written to disk in a streaming fashion.
Hence, for a stack of $n$ filter with size $s_j$ in the $z$-dimension, the number of whole volume reads and writes is reduced to 1 from $n$, at the cost of moderately increasing the simultaneously required main memory usage from $\underset{j \in \{1 \ldots n\}}{\max} s_j$ to $\underset{j \in \{1 \ldots n\}}{\sum} s_j$.
The application of a stack of two filters is illustrated in \autoref{fig:filter_stack}.

\subsection{Rendering}
For rendering of out-of-core volumes, Voreen features an octree-based multi-resolution volume raycasting renderer~\cite{brix2014visualization} that is capable of rendering multi-channel volumes.
Such multi-channel volumes often occur in the biomedical context, for example, when imaging a specimen multiple times with multiple fluorosphores.
Due to small errors in the imaging process, the resulting channels of the volume may be shifted slightly with respect to one another, which can be compensated for in the rendering process without resampling of the whole volume.
Voreen's interactive, progressive rendering scheme itself is immediate when the camera is moving in the scene, i.e., in each frame a new image is rendered using bricks of the octree structure at the desired resolution that are available in the VRAM of the graphics card.
When the camera is stationary, first an immediate frame with the maximum possible resolution is rendered, similar to the moving camera case.
Then the view is updated and refined incrementally by streaming required octree bricks from RAM (or disk, if required).
A brick ultimately updates the buffers of all rays that intersect it and updates the rendering displayed to the user depending on the compositing mode:
For maximum intensity/opacity (MIP/MOP) projection, the rendered image can be updated immediately if an updated intensity/opacity value exceeds that of the previous rendering, while for direct volume rendering (DVR) pixels are only updated when all bricks intersecting the corresponding ray have been processed.
This progressive, streaming rendering procedure is shown in action in \autoref{fig:progressive} where bricks on the left side of the image have already been rendered using the specified maximum resolution, but the right part is still displayed with lower fidelity.

Simple 2D slice views of large volumes are, of course, also supported (in $xy$-direction even without construction of an octree representation).
Out-of-core slice rendering for orientations other than in the $xy$-direction is done using the octree representation, by loading bricks that intersect the current slice.
If necessary, a lower resolution version is rendered during interaction before refining the rendering to the selected full resolution afterwards.

\subsection{Interactive Segmentation}
For arbitrary interactive segmentation of volume regions, an implementation of the random-walker method~\cite{grady2006random,wang2019review} was added to Voreen in conjunction with an uncertainty visualization~\cite{prassni2010uncertainty} already in 2010.

Recently, a hierarchical framework that allows the random walker method to be applied to very large out-of-core volumes~\cite{drees2021hierarchical} was developed and integrated into Voreen which in effect allows the interactive system by Pra{\ss}ni et al.~\cite{prassni2010uncertainty} to be applied to arbitrarily large volumes.
There, the octree representation of an input volume is used by applying the regular random walker method to bricks of the octree top-down, from the coarsest to finest resolution level.
Existing output foreground probability maps from the coarser layers are used to provide continuous seeds (in addition to the user provided once for the current brick, if present) which provide a global context and are used to refine the segmentation in the current brick, using a higher voxel resolution.
The method is made interactive by heavily pruning the resulting probability map octree during construction by stopping the refinement if bricks are homogeneous, definitely part of background/foreground.
If an output brick is similar to the previous result, an entire branch of the previous result can be reused.
Response times to individual edits of the label set are in the order of seconds to one minute in typical cases.
This is also demonstrated in the case study in~\autoref{s:eval_large_img}.
While the application is updating the foreground segmentation, the user can continue updating the label set, enqueueing another update operation to start immediately after the first one finishes, thus enabling a seamless interactive workflow.

\subsection{Vessel Network Analysis}
Voreen incorporates several methods for processing and analysis of vessel networks embedded in volumetric images.
The multi-scale vessel enhancement filtering method by Sato et al.~\cite{sato1997threed} is implemented as a multi-pass out-of-core procedure on volumes.
For a specific scale, the procedure consists of the application of a volume filter (see \autoref{sec:filtering}) to first extract the eigenvalues of the Hessian matrix which are used to compute the scale-dependent per-voxel vesselness.
For each scale, the generated slices are used to update the on-disk output volume which ultimately holds the maximum vesselness over all scales.

Volumes that have been converted into a binary foreground/background map of the vessel network (e.g. by thresholding preceded by vessel enhancement filtering, by interactive random walker segmentation, or by other, external means) a symbolic (i.e., graph) description of the vessel network can be extracted using a recently developed scalable, robust, and unbiased procedure~\cite{drees2021scalable}.
In addition to basic connectivity information, the extracted graph includes a number of morphological and geometric properties (such as length, straightness, volume, average radius, average roundness) for each edge (representing a vessel segment between branching points), as well as the centerline.
The implementation consists of a multi-stage pipeline that is evaluated iteratively.
Each stage has been engineered to require a maximum of $\bigO(m^\frac{2}{3})$ memory for an input volume of $m$ voxels.
This is achieved by either processing the volume slice-by-slice in a single sweep over the volume or by mapping the volume to memory using operating system capabilities and ensuring memory locality during access.
Other data structures such as intermediate graph representations or $k$d-trees are constructed by writing them as append-only binary representations to disk and later accessed by mapping the resulting files to memory.

After extraction, methods for quantitative comparison of vessel graphs, both in terms of topological~\cite{drees2019gerome} and geometrical similarity~\cite{mayerich2012netmets} between vessel segment centerlines, can be applied.
Application are, for example, a comparison of specimen or a performance analysis of segmentation algorithms.

\subsection{Quantification}
Volumes can be compared using voxel-wise measures such as the average intensity difference or (assuming binary input volumes) the Dice score.
This is done by scanning over both volumes simultaneously and comparing voxel intensities in corresponding slices.
For quantification of large volumes (or further processing), an implementation of the streaming connected component analysis by Isenburg and Shewchuk~\cite{isenburg2009streaming} (including support for export of generated metadata) is included.
The implementation scans over the volume twice, creating a \textit{root file} in the first scan, which is converted to a \textit{merge file} and used in the second scan to assign the final IDs to all identified components.
The cell cluster splitting method of Scherzinger et al.~\cite{scherzinger2016automated} can be used to quantify the number of nuclei in clusters in ultramicroscopy images.
It can operate on large volumes by first identifying clusters using the connected component finding procedure~\cite{isenburg2009streaming} and loading individual clusters from the disk representation for the splitting step.

\section{Visual Ensemble Analysis}
\label{s:visualensembleanalysis}

\begin{figure}
    \centering
    \includegraphics[width=\columnwidth]{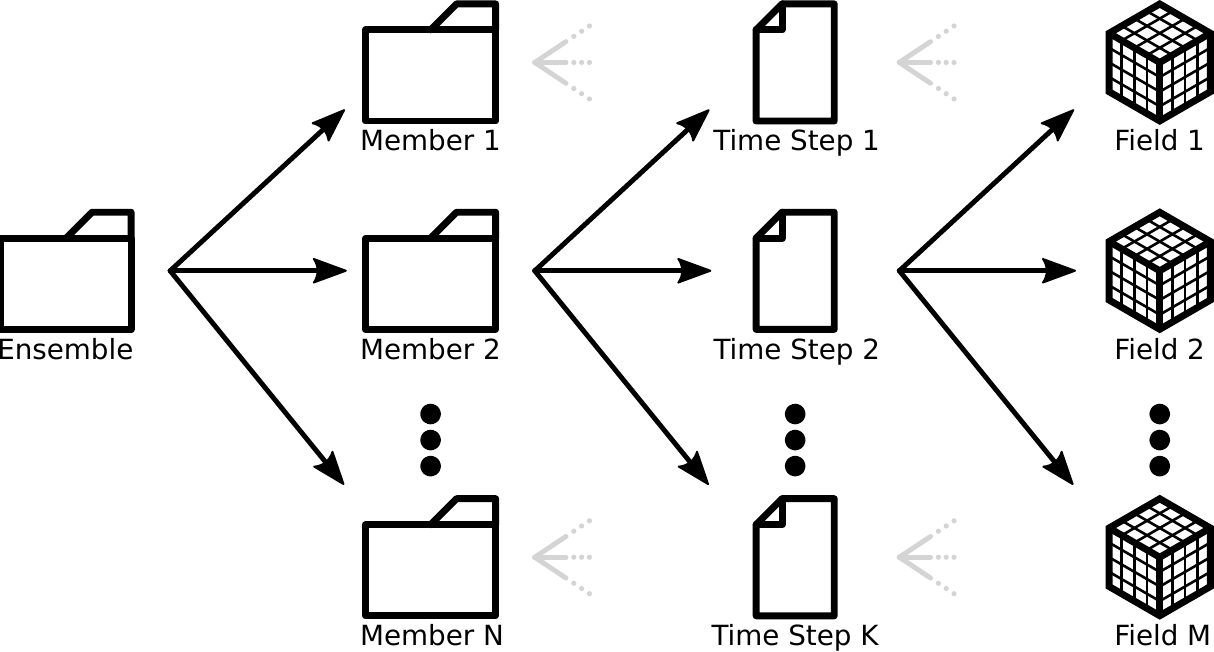}
    \caption{A schematic depiction of the ensemble data structure on file system level and its implementation in the Voreen framework. }
    \label{fig:ensemble}
\end{figure}

Many tasks involve the generation of ensemble data sets, such as fluid simulation analysis or cohort studies in medicine. 
Typically, such ensembles involve dozens of members each of which provide a time series of possibly multiple scalar or vector fields (cf.~\autoref{fig:ensemble}).
These so-called spatio-temporal multi-run multi-field ensemble data sets need to be visualized in order to be understood by domain experts.
Not every user or institution has access to compute cluster resources (sometimes due to data privacy issues) that would allow for distributed processing of the data. Hence, the development of visualizations, algorithms, and strategies is required that can handle the data on consumer hardware.

\subsection{Ensemble Data Access}

To allow for large ensemble data sets to be processed and rendered interactively, an ensemble data structure is set up in the first and initial loading step. Hereby each volume file is once loaded into main memory to calculate per-field minimum and maximum values, if not already present as part of the volume file meta data. In the same step, also book-keeping of meta data such as union and intersection of the spatial and temporal domains is performed. These meta data are then stored to disk using a json file.  Once another loading attempt is made, the json file is loaded instead of the ensemble volume data itself. Subsequent processing and rendering steps will then only load the required volume files on demand. Typically, this is performed asynchronously as it is for most implemented algorithms.
Stackable filter operations are implemented that allow us to filter the data set with respect to member names, time range, and field modality.

To generate an ensemble to be loaded by Voreen, it simply is required to put respective volume files (of any supported format) into a directory tree depicted in \autoref{fig:ensemble}. The root folder contains a separate directory for each ensemble member where the name of the directory will be used, e.g., in the UI as name for the respective member. Since directory names must differ on file system level, so will member names, which therefore are treated as unique identifiers. Each member directory may contain multiple volume files, one for each time step. The names, e.g., when sorted lexicographically as it is common for numerical simulation data, might indicate the order of the time steps. However, file format specific meta data, if available, is also considered to sort the time steps correctly. Each volume may have unique dimensions, spacing, offset and transformation, as well as time stamp and even data type. Additionally, the available fields across members or time steps do not need to match each other. Most implemented algorithms only make use of the common set of fields, but accessing the non-common fields still is possible. 
To store multiple field modalities within each file, a respective format has to be chosen that supports multiple volumes and meta data such as VTI, NetCDF-CF, or HDF5. If the ensemble only contains a single field, any supported volume format can be used.

In contrast to large volume data sets, we define large ensembles to be a larger set of volumes  that fit in main memory individually, but not as a whole.
To handle the amount of data, implementing an on-demand reloading mechanism for individual volumes is a necessity. We want to make sure that the memory management is transparent to the developer and, of course, also to the end user. To solve this, we only load relevant meta data of the volume files such as its dimensions, spacing, and transformation, if supported by the format. As soon as the actual volume data are requested, the entire data will be read from disk into main memory. As the main memory fills quickly, e.g., when loading large time series, volumes are removed from main memory according to the Least Recently Used (LRU) strategy. 
If loading meta data separately from the volume data is not supported for a format, a good option is to convert the files using Voreen's bulk processing capabilities. We recommend to make use of the HDF5 format as it supports multiple fields per file as well as data compression.

\subsection{Multi-run Similarity Plot}

Members of ensemble data sets typically differ by their initial conditions and parameters in case of simulated data, or treatment, when cohort studies are considered. 
A typical scenario therefore is to analyze the influence of the respective setting, especially on the evolution over time (for temporal data). Investigating all volumes of time-varying data for many ensemble members quickly becomes an overwhelming endeavor. Instead, to reduce the  cognitive load during the analysis, it can be advantageous to visually encode the entire ensemble and its evolution over time using data abstraction and visual summarization. 
To generate such a single overview plot, we transform the set of volumes to a similarity space.
Here, we make use of the multi-run similarity plots proposed by Fofonov et al.~\cite{fofonov2019projected}.
The idea is to create a lower-dimensional embedding of the whole ensemble in which each time step of each ensemble member is represented by a single point in the embedding such that  Euclidean distances of the points in the embedding represent dissimilarities of the respective fields. The interpretation of the Euclidean distances was shown to be intuitive to experts from different domains~\cite{leistikow2020interactive}.

To create such a lower-dimensional embedding of the similarity space, a distance matrix is required that encodes pairwise dissimilarities between individual time steps. A distance metric is required that allows to calculate each entry, i.e., the dissimilarity between two respective volumes.
Iso-surfaces were shown to be good scalar field descriptors and can therefore be used for this purpose, but do not capture the entire field. Fofonov et al.~\cite{fofonov2019projected} generalized the idea of iso-surface similarity to a field similarity that was proven to have more desirable results to create embeddings than other metrics based on gradients or correlation. For this reason, we decided to integrate the scalar field similarity by Fofonov et al., which is used by default. If desired, any other scalar field similarity measure could quickly be added using Voreen's architecture.
We further extend the approach to a multi-field similarity metric that allows us to also calculate similarity between vector fields. Vector field similarity is achieved by separating a vector field into its magnitude and direction component, computing the respective similarities, and combining the resulting distance metrics into a new distance metric~\cite{leistikow2020interactive}.
The resulting pairwise dissimilarities are stored in a distance matrix.

Since each entry of such a matrix encodes the dissimilarity between a particular pair of time steps and a metric is used to calculate the dissimilarity, the matrix turns out to be symmetric which allows us to only store, e.g., the upper triangular part to reconstruct the matrix.
In principle, generating a distance matrix for a single field only requires two volumes to be stored in main memory at the same time. However, this would require us to load the same volumes multiple times which slows down the process since disk IO is the main bottleneck. To remedy this problem and, to reduce the overall calculation time, we first load each volume once and apply a Monte-Carlo sampling. The samples are only distributed within the common spatial domain which can be further restricted by providing a sample mask, and are stored in a per-field feature vector that is written to disk by using a memory mapped file. This allows even large ensembles to be processed efficiently, since parts of the file that are written to and read from will reside in main memory. The operating system is hereby responsible for deciding how exactly the mapping is performed. Another advantage is that the calculation can be aborted in between and restarting will pick up the memory mapped files that have already been created up to the point of interruption. As soon as all feature vectors are available, the distance metric can be applied in parallel on multiple time steps at the same time, depending on the available computing resources. Since the calculation of the distance matrix scales quadratically in the number of time steps and can therefore take a decent amount of time (cf.~\ref{s:eval_large_ensemble}), we consider it as preprocessing step that could be executed in a headless environment. Still, the calculation is performed asynchronously and could therefore also be integrated directly into a more complex, visual application. 

\begin{figure}
  \includegraphics[width=\columnwidth]{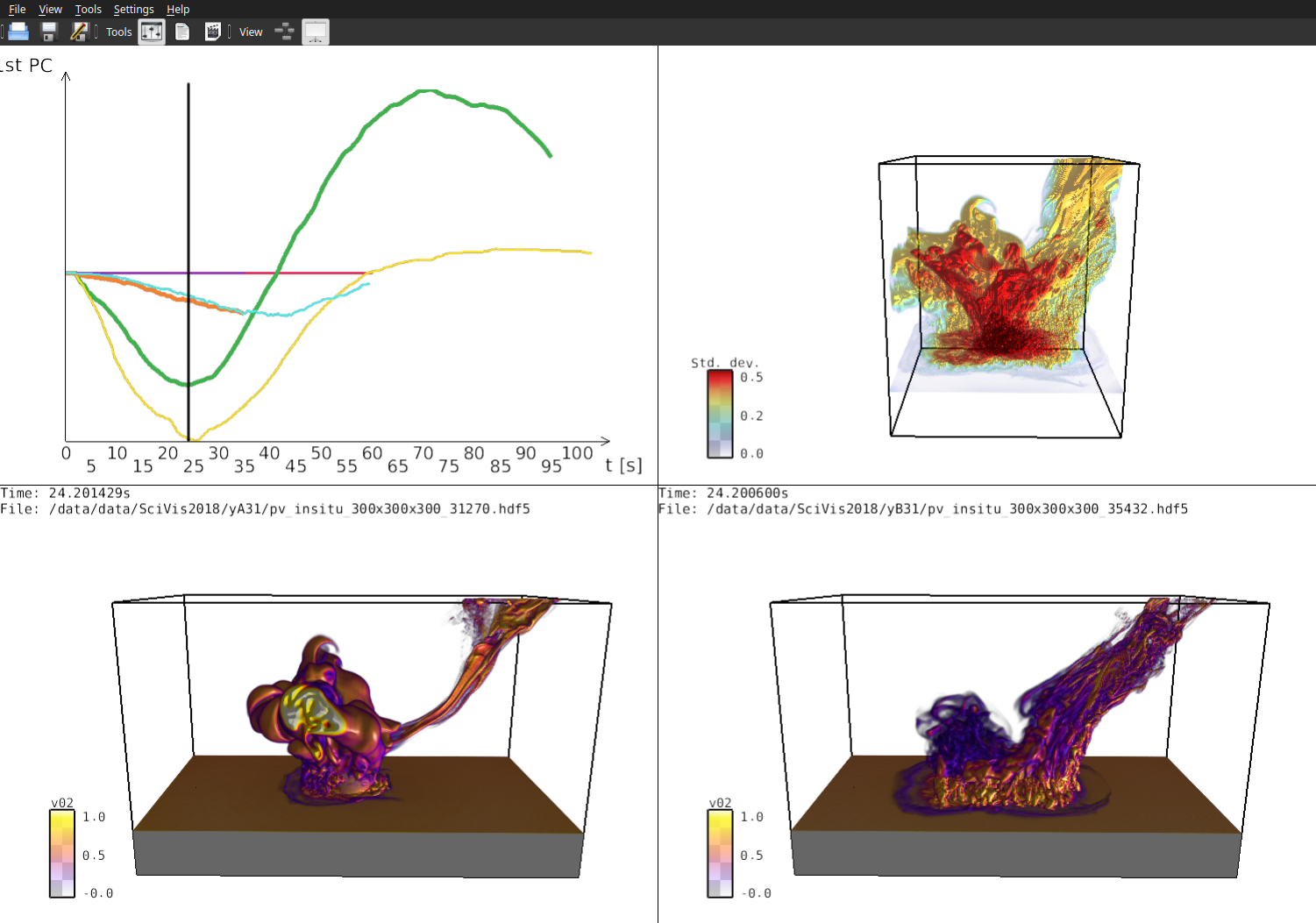}
  \caption{
    A screenshot of the Voreen application mode interface with multiple linked views providing insights in the \SI{840}{\giga\byte} deep water asteroid impact ensemble data set~\cite{asteroiddata}. It features a multi-run similarity plot (upper left) as proposed by Fofonov et al.~\cite{fofonov2019projected}, the ensemble variance at the selected point in time (upper right), and two juxtaposed volume renderings of two individual simulation runs (bottom). The analyzed scalar field is the volume fraction of water.
  }
  \label{fig:ensemble-similarity}
\end{figure}

A classical Multi-dimensional Scaling~\cite{wickelmaier2003introduction} approach is then used to project the distance matrix into a lower-dimensional space.
Points of consecutive time steps can be connected to form curves in the embedding (see~\autoref{fig:ensemble-similarity}).
The number of dimensions to be used for the embedding depends on the intrinsic dimensionality of the data and can be assessed, e.g., by the use of bar charts depicting the principle components. In most of the cases, up to three dimensions were sufficient for the data sets we used so far. Hence, either the first principle component can be visualized using time as secondary axis, or two and three principle components are used as axes, respectively, visually encoding the temporal information by using a linear color gradient along the generated curves.  The time required to calculate the embedding from a distance matrix depends on the number of time steps and fields but usually is finished in the order of a second (cf.~\ref{s:eval_large_ensemble}).

We further implemented interaction mechanisms to allow for making use of the similarity plot in a more complex application. As such, two separate selections can be performed by clicking anywhere on the curves. A typical use case would be to select two simulation runs that are either quite similar or dissimilar (i.e., either close or far in the projected space) at a particular point in time and to compare their volume renderings in juxtaposition (cf.~\autoref{fig:ensemble-similarity}).
The selection of the temporal domain might also be used to calculate and render the respective ensemble mean and variance (by means of standard deviation)~\cite{leistikow2020interactive}.
Moreover, a sub-selection of members can be performed interactively to generate a new embedding only considering the selected data volumes.

\subsection{Parallel Coordinates}

Parallel Coordinates~\cite{Inselberg1996ParallelCA} are a common tool that allow for finding correlations between values of scalar fields of multi-field or multi-channel data.
As the multi-run similarity plot allows us to analyze intra-field similarity, parallel coordinates are an orthogonal extension, which we integrated into our framework. 
We again make use of Monte-Carlo sampling on the (optionally masked) spatial domain.
For fields with multiple channels, e.g. vector fields, each channel is treated as a separate axis.
On the temporal domain a uniform sampling is applied.
All samples are then stored in a linear array that can be stored to disk, combined with information on how the data shall be interpreted. For example, offsets are stored to address individual runs and field axes. 
We then upload the data directly to the GPU to efficiently render the parallel coordinates plot.
Commonly applied interaction mechanisms are supported, i.e., we support pairwise swapping of axes as well as brushing on individual axes.
Overlapping lines may be rendered using density-based blending.
For up to four axes, a transfer function can be defined that can be applied in a linked multi-volume rendering.
Each selected section will then clamp the respective section on the respective transfer function.
The selection updates the respective renderings immediately, as only an update of the transfer function is required, which is applied on the GPU.
Additionally, the intersection of all axes is calculated and can be linked to another spatial volume visualization where only those voxels are included that satisfy the brushing selections on all axes. This process needs more processing time, as it requires us to iterate over all involved voxels to test for an intersection. It is therefore implemented asynchronously to maintain a responsive user experience.
Finally, the parallel coordinates can also be exploited as time histogram for one selected field, i.e., each axis represents one time step of the field in chronological order.

\section{Evaluation}
\label{s:evaluation}

To demonstrate how out-of-core processing and rendering of both large imaging and ensemble data can be performed in the context of interdisciplinary research, we provide two case studies.

\subsection{Large Imaging Data}
\label{s:eval_large_img}

To demonstrate the capabilities in processing and rendering of large single volume data sets, we segmented and analyzed the vessel tree in a light sheet microscopy scan of a mouse.
The results presented here

were obtained on a midrange consumer PC with an Intel i5-6500 (\SI{3.2}{\giga\hertz}, 4 cores), \SI{16}{\giga\byte} RAM, a single NVidia GTX 1060 (\SI{6}{\giga\byte} VRAM) and a \SI{1}{\tera\byte} Samsung EVO 970 solid state drive.

The original data set occupies roughly \SI{377}{\giga\byte} of disk space which corresponds to $9,070\times 12,723 \times 1,634$ voxels at $16$ bit per voxel.
The voxel spacing is $\SI{750}{\nano\meter}\times \SI{750}{\nano\meter}\times \SI{3}{\micro\meter}$ which results in a real world size of $6.8 \times 9.6\times \SI{4.9}{\milli\meter\cubed}$.
Due to the limited available space on the internal solid state drive, the original data set was read from an external hard disk to create an octree representation of the data set, which was automatically stored and cached on the internal drive.
This process took \SI{66}{\min} and was primarily limited by the speed of the hard drive (roughly 60 to $\SI{120}{\mega\byte\per\second}$).

The arterial vessel tree was then segmented semi-interactively up to a vessel radius of about \SI{50}{\micro\metre} using the hierarchical random walker framework~\cite{drees2021hierarchical}.
This process required roughly \SI{3}{\hour}.
Isolated updates to preliminary segmentation by edits to the label set were computed in anywhere between \SI{1}{\second} and \SI{1.5}{\min}, depending on the size of the labeled structure, other load on the machine, and whether the affected portions of the volume are cached in memory due to recent access.
In practice, while the segmentation update is computed, the user places multiple additional foreground and background labels in the vicinity of previous edits, either to correct previous mislabelings, or to label additional structures in the image.
While processing a larger number of edits to the label simultaneously set may increase the computation time in comparison to a single edit, the relationship between number of edits and computation time is sublinear, since the sets of bricks that require updates typically overlaps between label edits, especially in coarser levels of the tree, or if labels are spatially close.
The total update time was always below the time required to recompute the segmentation from the full set of labels without a previous solution, which was \SI{11}{\min} in this case.
Closing the application between labeling sessions is unproblematic because the segmentation can be loaded from the previous solution within seconds (equivalent to a segmentation update without changes to the label set), which can be stored in a fixed location on disk, and since the octree representation of the input data set is cached by Voreen as well, as described above.
A 3D raycasting rendering of the original data set and the foreground segmentation in the task-specific labeling application is shown in \autoref{fig:progressive}.

After the semi-automatic segmentation of the vessel lumen, a graph representation (with end points and branching points as nodes and connecting vessel segments as edges) is created.
Before the graph extraction, the segmentation is postprocessed by surface smoothing using a binary median filter of size $11\times11\times3$ and by removal of cavities (i.e., disconnected smaller components of the background) and small foreground objects using the streaming connected component analysis implementation~\cite{isenburg2009streaming}.
The surface smoothing and removal of background and foreground components required \SI{148}{\minute}, \SI{129}{\minute}, and \SI{123}{\minute}, respectively, while the vessel graph extraction finished after 7 iterations, after roughly \SI{31}{\hour}.
These computations were performed in the graphical application here (which allows, for example, the visualization of intermediate results while the computation is still ongoing), but can instead also be performed headlessly, for example on a dedicated server.

The resulting vessel graph (including centerlines and per-edge/vessel segment properties) can then be saved to disk, either including full centerline information as a json file or as aggregated properties for edges and nodes in a csv file.
Additionally, the graph structure can be visualized in conjunction with the volume data.

The built-in animation editor can then be used to generate video material, showcasing the research to colleagues or for outreach.
As an example, camera positions can be set as fixed animation keys during which intermediate camera positions can be interpolated resulting in a smooth camera track.

\subsection{Ensemble Data}
\label{s:eval_large_ensemble}

To demonstrate the ensemble processing and visualization capabilities, we walk through an analysis workflow for the deep water asteroid impact ensemble data set~\cite{asteroiddata, leistikow2019aggregated}. The simulated asteroid in each of the 7 simulation  runs (i.e., 7 ensemble members) differs in size, airburst height (if any), and approaching angle (i.e., 3 simulation parameters). The considered simulation output contains a pressure and temperature field, as well as volume fraction fields of water and asteroid. All fields are stored into volumes of $300^3$ voxels at 32 bit per voxel resulting in $\SI{108}{\mega\byte}$ per volume. The spatial domain of the runs mostly match each other, except for a single run covering a slightly smaller region as here the impact angle was steeper when compared to the other runs. In total, 1,945 time steps are available, unevenly distributed between the runs. This adds up to \SI{840}{\giga\byte} of data, which we converted into the HDF5 file format to make use of compression.
For this case study, a similar hardware configuration was used, i.e. an Intel i7-6700k CPU (\SI{4}{\giga\hertz}, 4 cores), \SI{16}{\giga\byte} RAM, \SI{1}{\tera\byte} SanDisk PLUS SSD, and an NVidia GTX 1080 graphics card (\SI{8}{\giga\byte} VRAM).

In a one-time preprocessing step, the ensemble data were loaded into Voreen for the first time. Since only relevant meta data need to be loaded from disk, this only required $\SI{15}{\second}$. Next, we computed the similarity matrices for each field using 8,192 spatial samples which took around $\SI{56}{\minute}$ for the entire ensemble. 
Before we calculated the required data for the parallel coordinates visualization, we restricted the sampling to only consider the common spatial domain of all runs and used 16,384 spatial and $100$ temporal samples. After roughly $\SI{13}{\minute}$ the calculation finished and we stored both the similarity matrices and the parallel coordinates data to disk. The calculations could also be performed headless, e.g. on a dedicated server. However, as the calculation are performed asynchronously and were expected to finish within a reasonable amount of time, we calculated them from within the visual application.

Consequently, we switched to an analysis workspace in application mode comprising multiple linked views and loaded both the ensemble and preprocessed data.
The similarity plot for all members and fields was calculated automatically within roughly \SI{3}{\second} as soon as the required ensemble data and similarity matrices were loaded. 
For the analysis, an interactive experience is essential. The only data that need to be loaded from disk at this point are the individual volumes, if the user requests the data (i.e., on demand), e.g., when performing a selection of a particular time step for a volume rendering. This means that the bottleneck again is the sequential read speed of the disk where the ensemble is stored.
Depending on the amount of RAM and the size of an individual volumes, a couple of them are cached once they were loaded into main memory. However, caching virtually has no advantage if selections are performed in random access on the temporal domain and across fields. As our drive has sequential read speeds of up to $\SI{535}{\mega\byte\per\second}$, loading a non-cached volume still only took much less than a second. Caching of volumes instead becomes useful if a smaller time interval is considered multiple times, e.g. when creating an animation on volume renderings.
The same can be applied to parallel coordinates. Brushing on the axis only affects the respective transfer function which can then be applied to any volume of the respective field on the GPU.
Hence, just the respective volume file needs to be loaded from disk, in case it does not yet reside in main memory.

\section{Limitations}
\label{s:limitations}

The proposed framework is capable of handling large data in the form of single volumes and ensembles. However, the proposed algorithms are still limited by the hardware that is used. In the following, these limitations will be discussed.

The presented slice-based filter stack operates on slices, hence it is required that each slice fits into main memory. Depending of the filters being used, even multiple slices need to be kept in main memory, at least one input and one output slice at a time. As an example, a volume of dimensions $10000^3$ (\SI{4}{\tera\byte}) can be split into $10000$ slices of size $10000^2$ which would still require $10000^2 \times \SI{4}{\byte} \times 2 = \SI{800}{\mega\byte}$ of main memory, assuming each voxel value is represented by 4 Bytes.

The octree raycaster has 25 bits available to address nodes of the octree, which means that trees to be rendered must comprise of less than $2^{25}$ nodes.
For a brick size of $32 \times 32 \times 32$ this means that the maximum volume size for rendering is (at least, depending on the exact volume dimensions) $\SI{1}{\tera\byte}$.
Larger bricks allow for rendering of larger data sets (e.g., at least $\SI{8}{\tera\byte}$ for bricks of $64^3$ voxels).

For ensemble data at least a single field of a single time step of a single member needs to fit into main memory at a time in the current implementation. However, this is not a problem currently, since, for example, volumes in the deep water asteroid impact data set~\cite{asteroiddata}, are far from reaching the limitations of main memory (\SI{108}{\mega\byte} per volume). Furthermore, this is no principle limitation, as it is feasible to also make use of the octree infrastructure for ensembles in the future. The more individual fields fit into main memory, the more responsive the overall interactive experience will be, as fields that are no longer resident in main memory need to be reloaded on demand. It is therefore still recommended to store the data on fast mass storage (e.g., solid state drives) to enable an interactive in-detail analysis on large ensemble data sets. For data aggregating algorithms such as the one underlying the Similarity Plot, this is not a necessity, since only the visualization on the already aggregated result is required to be interactive.

Voreen is designed to be a \textit{single user} workstation application and as such does not support distributed computation, e.g., on compute clusters.
Additionally, all GPU accelerated operations currently do make use of additional graphics cards installed in the machine, which, however, may be changed in the future.

\section{Conclusion and Future Work}
\label{s:conclusion}
We have presented recent advancements to the rapid prototyping framework Voreen. 
These include various features that enable and enhance interdisciplinary research as well as dedicated support for out-of-core single volumes as well es spatio-temporal multi-run multi-field or multi-channel ensembles.

In the future we would like to further expand Voreen's feature set with regards to handling out-of-core data sets.
For example, we would like to make use of the octree infrastructure for ensembles data sets.
We will research how other existing image analysis and processing methods can be generalized for out-of-core application and integrate them into Voreen.
In particular, machine learning approaches seem to be suitable for application on the LOD octree datastructure.
We also plan to address the comparative analysis of cohorts to further establish Voreen for interdisciplinary research.

\section*{Acknowledgments}
This work was funded by the Deutsche Forschungsgemeinschaft (DFG) – CRC 1450 – 431460824.

\bibliographystyle{abbrv-doi}

\bibliography{references}
\end{document}